\documentclass[sigconf]{acmart}

\AtBeginDocument{%
  \providecommand\BibTeX{{%
    \normalfont B\kern-0.5em{\scshape i\kern-0.25em b}\kern-0.8em\TeX}}}

\copyrightyear{2023}
\setcopyright{none}
\acmConference[Conference Acronym]{Conference Acronym}{2023}{Preprint}

\usepackage{graphicx}
\usepackage{subcaption}
\usepackage{multirow}
\usepackage{tabularx}
\usepackage{xcolor}
\usepackage{balance}

\begin{document}

\title{A Comparative Study of Reference Reliability in Multiple Language Editions of Wikipedia}

\author{Aitolkyn Baigutanova}
\orcid{0000-0002-9852-4157}
\affiliation{
  \institution{KAIST, IBS}
  \city{Daejeon}
  \country{South Korea}
}
\email{aitolkyn.b@kaist.ac.kr}

\author{Diego Saez-Trumper}
\orcid{0000-0002-7679-5423}
\affiliation{%
  \institution{Wikimedia Foundation}
  \city{Barcelona}
  \country{Spain}
}
\email{diego@wikimedia.org}

\author{Miriam Redi}
\orcid{0000-0002-0581-0251}
\affiliation{
  \institution{Wikimedia Foundation}
  \city{London}
  \country{United Kingdom}
}
\email{mredi@wikimedia.org}

\author{Meeyoung Cha}
\orcid{0000-0003-4085-9648}
\affiliation{%
 \institution{IBS, KAIST}
 \city{Daejeon}
 \country{South Korea}
 }
 \email{meeyoungcha@kaist.ac.kr}

\author{Pablo Aragón}
\orcid{0000-0002-6017-4577}
\affiliation{%
  \institution{Wikimedia Foundation}
  \city{Barcelona}
  \country{Spain}
}
\email{paragon@wikimedia.org}




\begin{abstract}
Information presented in Wikipedia articles must be attributable to reliable published sources in the form of references. This study examines over 5 million Wikipedia articles to assess the reliability of references in multiple language editions. We quantify the cross-lingual patterns of the \textit{perennial sources list}, a collection of reliability labels for web domains identified and collaboratively agreed upon by Wikipedia editors. We discover that some sources (or web domains) deemed untrustworthy in one language (i.e., English) continue to appear in articles in other languages. This trend is especially evident with sources tailored for smaller communities. Furthermore, non-authoritative sources found in the English version of a page tend to persist in other language versions of that page. We finally present a case study on the Chinese, Russian, and Swedish Wikipedias to demonstrate a discrepancy in reference reliability across cultures. Our finding highlights future challenges in coordinating global knowledge on source reliability.

\end{abstract}



\maketitle

\section{Introduction}

As a global knowledge encyclopedia, Wikipedia is one of the most popular websites worldwide. It is open access and maintained by the online community that collaboratively creates content following specific editing policies. A core content policy is verifiability~\cite{verifiability}, which requires that information included in Wikipedia articles is supported by reliable and relevant references from which claims are derived. In order to encourage editors to avoid untrustworthy references, the Wikipedia community created a reliability index called the \textit{perennial sources list}~\cite{PerennialSource}. This list contains web domains and their labels (e.g., blacklisted, generally reliable), assigned based on a collective consensus through discussions among the editors~\cite{noticeboard}. Recent research has shown that these community efforts have led to improved reference quality in English Wikipedia~\cite{ours}.

A growing body of research explores the role of references in the encyclopedia. A study of sources cited in scientific articles has shown that Wikipedia relies heavily on prestigious journals from STEM fields, with sources from non-STEM disciplines being marginal but relevant in biographical content~\cite{yang2022map}. Recent research also found a moderate yet systematic liberal polarization in the selection of media and news sources~\cite{yang2022polarization}. While these studies are limited to English Wikipedia, there exist a few works that examined the popularity of sources across different languages of Wikipedia~\cite{lewoniewski17, lewoniewski23}, finding that some sources are common across editions but that each language often has its own distinct set of sources. However, the reliability of sources across a broader set of language editions of Wikipedia remains unexplored.

Understanding referencing behavior across languages is critical in multilingual ecosystems such as Wikipedia. The English Wikipedia edition is currently the largest in terms of community size and the number of articles, but it only covers a portion of the information found on other Wikipedias~\cite{hecht2013mining}. The impact of this dominant language edition on other editions remains unclear~\cite{valentim2021tracking}, and the cross-lingual effect on Wiki-related policies is still being discussed. One study has shown that even if large language editions share core editing rules, localized rule sets tend to become increasingly diverse over time~\cite{Hwang_Shaw_2022}.
 
Here, we examine the reliability of references in over 300 language editions of Wikipedia to study the cross-edition effects on reference quality: how do the English editor community's initiatives affect the referencing behavior of other language editions? We investigate the reliability of the most common sources and their presence in the same article across languages and find that non-authoritative references in the English edition tend to persist in other language versions of the same articles. We present our preliminary findings on the potential risks to content verifiability that may result from translating articles into less developed language editions.

While we hypothesized that a community-maintained list of untrustworthy sources in one language (i.e., English) can be a good starting point for improving reference quality in other smaller language editions, our data indicated that some domains deemed untrustworthy and hence banned in the English edition are still actively used in some language editions. One example is the Russian language online newspaper \textit{lenta.ru}, which frequently appears in Russian Wikipedia.The daily tabloid \textit{huanqui.com} which is under the auspices of the Chinese Communist Party, was another example that frequently appeared in Chinese Wikipedia. These web sources likely provide cultural- or regime-specific information and values that are not universally shared by all cultures. Wikipedia, however, is read by a global audience and serves as a foundation for many large language models and search engines. Therefore, our discovery of a discrepancy in reference reliability across cultures opens up a discussion about the need to account for the broad audience of this interconnected world when deciding ``global knowledge''.

\section{Data}\label{sec:data}
\subsection{Reliability of Sources in Wikipedia}

We use the internal labeling of source reliability decided by the Wikipedia community, called the \textit{perennial sources list}. It comprises a collection of web domains in five categories: (1) blacklisted, (2) deprecated, (3) generally unreliable, (4) no consensus, and (5) generally reliable. Following the methodology of a recent study~\cite{ours}, we refer to the first two categories as non-authoritative sources. 
As of May 2023, a \textit{perennial sources list} page exists in the following 12 language editions of Wikipedia: English, Russian, French, Persian, Swedish, Chinese, Portuguese, Greek, Lithuanian, Turkish, Vietnamese, and Nepali. Only the first six are actively maintained, with no updates made to the classification of the remaining lists this year. Furthermore, the Persian Wikipedia list includes only 14 web sources, while sources included in the French classification do not have an explicit reliability label. 

For the remaining four lists in English, Chinese, Swedish, and Russian Wikipedia, we show descriptive statistics in Table~\ref{tab:perennials}, including the number of classified domains, the overlap of the sources in the list with English Wikipedia's \textit{perennial sources list}, and the coverage by the lists of the corresponding language edition and English edition. We define coverage as the percentage of pages in the corresponding language edition with at least one citation to the sources from a \textit{perennial sources list}.

\begin{table}[t]
\caption{Descriptive statistics of \textit{perennial source lists} in four language editions. Coverage represents the percentage of articles citing at least one source from the corresponding list.}
\label{tab:perennials}
\resizebox{\linewidth}{!}{\begin{tabular}{c|c|c|c|c}
\hline
\textbf{\begin{tabular}[c]{@{}c@{}}Language\\ edition\end{tabular}} & \textbf{\begin{tabular}[c]{@{}c@{}}Number of\\domains\end{tabular}} & \textbf{\begin{tabular}[c]{@{}c@{}}Overlap with\\the English list\end{tabular}} & \textbf{\begin{tabular}[c]{@{}c@{}}Coverage\\ \end{tabular}} & \textbf{\begin{tabular}[c]{@{}c@{}}Coverage\\ (English list)\end{tabular}} \\ \hline
English                                                              & 1,156                                                                            & 100\%                                                                                & 22.9\%                                                                   & 22.9\%                                                                       \\
Chinese                                                              & 211                                                                             & 17.5\%                                                                               & 10.0\%                                                                   & 11.5\%                                                                       \\
Swedish                                                              & 51                                                                              & 54.9\%                                                                               & 0.8\%                                                                    & 2.4\%                                                                        \\
Russian  & 60  & 15.0\%  & 5.7\%     & 11.6\%    \\ 
\hline
\end{tabular}}
\end{table}

\subsection{Dataset Description}
We first retrieve the \textit{perennial sources list} from English, Chinese, Swedish, and Russian editions using Python's BeautifulSoup library to parse HTML documents. We use the online version of the lists from May 2023. Then, we manually filled in the missing data for the entries in the lists that did not include an explicit link to the source. This is the case for the Russian edition, where website domains are not listed for all the sources, but they can be inferred from the links to the discussion pages or the linked Wikipedia pages. As a result, we could obtain a table of sources along with their domain and category for each of the four editions. Only sources with explicit labels were included for the remainder of the analysis.

We used the following attributes from Wikidata dumps: item ID (a language-agnostic Wikidata page identifier), language edition ID, page ID in a given language edition, and the corresponding page title. We collect 2,1824,103 unique Wikidata items across all language editions. We consider the pages in the article namespace of Wikipedia. We combine this data with the XML dumps for the most recent Wikipedia page versions (as of February 2023) to retrieve the raw text of each article. Using this text, we enrich the dataset with sources included in the English \textit{perennial sources list} cited in a given article, along with the source's category. This results in 5,189,606 articles across 314 editions that include at least one reference to a \textit{perennial source}.


\section{Results}
\subsection{Reliability by Language Editions}
\begin{figure}[t!]
    \centering \includegraphics[width=.9\linewidth]{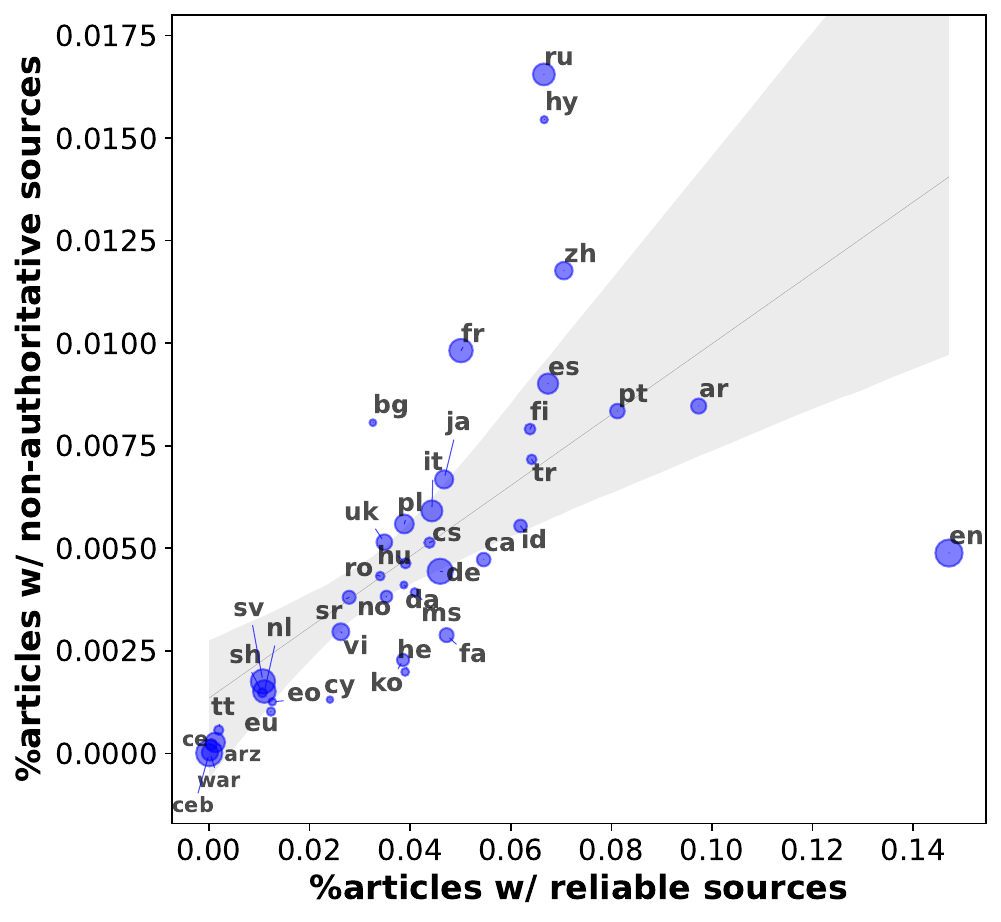}
     \vspace*{-3mm}
    \caption[]{The proportion of articles referencing reliable and non-authoritative sources in 40 language editions. The circle size indicates the number of articles, while the shaded area indicates a 90\% confidence interval.} 
    \label{fig:scatter50}
\end{figure}

To investigate the spread of English Wikipedia's \textit{perennial sources} across multiple language editions, we identify the proportion of articles in each edition that include at least one reference to these sources. Figure~\ref{fig:scatter50} shows the percentage of articles referencing reliable and non-authoritative sources in the 40 editions with the largest number of articles. Languages can be mapped from the Wikipedia code~\cite{list_of_wikipedias}.

The plot shows outliers in the two directions of the confidence interval represented by the gray area. On the one hand, the English edition is located below the confidence interval, meaning the proportion of articles citing reliable domains is larger. This observation is consistent with recent research~\cite{ours}, as the community of English Wikipedia is more aware of the non-authoritative domains listed in the local \textit{perennial sources list}. On the other hand, the outliers above the confidence interval appear to have a relatively larger proportion of articles citing deprecated or blacklisted domains. These are Russian (ru), Armenian (hy), Chinese (zh), French (fr), and Bulgarian (bg). We qualitatively assessed what non-authoritative sources are prevalent in these five editions. We found that for Russian and Armenian, the trend is primarily caused by the blacklisted domain \textit{lenta.ru}, a Russian online newspaper. For the French edition, the US-based source \textit{city-data.com} contributes the largest proportion. For the Chinese language edition, the trend is predominantly attributed to the Chinese tabloid  \textit{huanqiu.com}, and international media source \textit{epochtimes.com}. Finally, the deprecated British tabloid \textit{dailymail.co.uk} contributes a significant proportion of all the editions.

\begin{figure}[t!]
    \centering 
    \includegraphics[width=\linewidth]{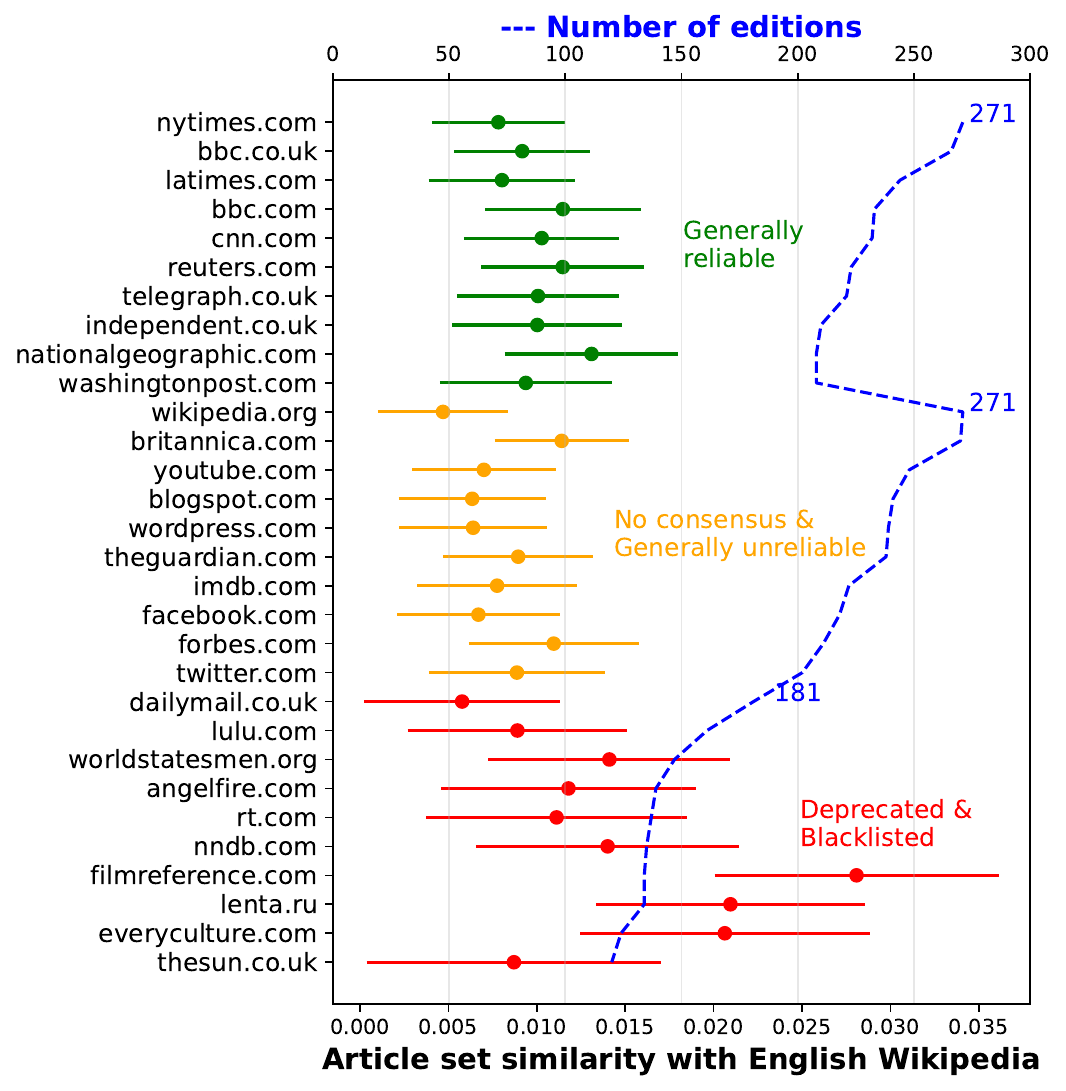}
    \caption{Average pairwise similarity of article sets citing the same source with English edition. The top ten most popular sources from each category are shown. The error bars indicate the standard error.} 
    \label{fig:setsim}
\end{figure}

We further examine specific domains that appear most frequently across multiple editions of Wikipedia. The blue dashed line in Figure~\ref{fig:setsim} presents the number of editions for each category's top ten most widely referenced domains. The most popular web domains across different languages are generally reliable \textit{nytimes.com} and \textit{bbc.co.uk}, generally unreliable \textit{wikipedia.org}, and no consensus \textit{britannica.com}, which are referenced in around 270 editions. We note that the most frequently used non-authoritative domain, namely \textit{dailymail.co.uk}, is ranked 34th, cited in 181 language editions. This indicates that although deprecated and blacklisted sources are less popular, they are still commonly cited.

\subsection{Reference Similarity with English Edition}

Next, we explore the relationship between references in English Wikipedia and the other language editions. We measure how likely sources are to appear in a non-English version of an article when they are also present in the English 
one. That is, we analyze pages present in at least two editions of Wikipedia, where one is in English. First, we gather a collection of item IDs from each edition, referencing a specific source. Then, we compute the pairwise Jaccard similarity coefficient of the set of items citing a given source in English and other language editions. This coefficient is a commonly used statistic to determine the similarity between two sets and is computed as a ratio of two sets' intersection over their union.

Sources in the deprecated and blacklisted categories of the \textit{perennial sources list} demonstrated a higher article set similarity with the English edition than generally reliable (t=6.08, p<0.001) and no consensus and generally unreliable (t=6.16, p<0.001) ones. In contrast, the generally reliable category also demonstrated a lower value on average than the middle two categories (t=2.62, p<0.01). Figure~\ref{fig:setsim} displays the average article set similarity of non-English editions with the English Wikipedia per source. The scores are shown for each category's ten most commonly referenced domains in the \textit{perennial sources list}. We observe that the highest similarity is exhibited by non-authoritative sources, such as \textit{filmreference.com}, \textit{lenta.ru}, and \textit{everyculture.com}. These findings may indicate that non-authoritative domains found in the English edition are more likely to be present in other language editions.

\begin{figure*}[t!]
\begin{subfigure}{.32\textwidth}
  \centering  \includegraphics[width=0.99\linewidth]{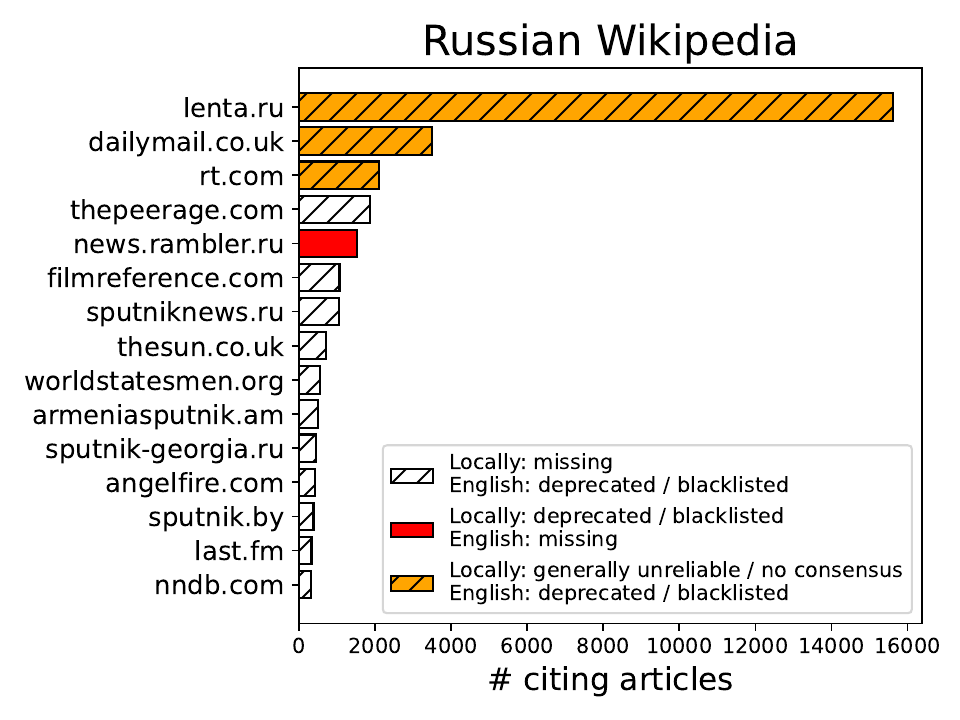}
  \label{fig:ru_barplot}
\end{subfigure}%
\begin{subfigure}{.32\textwidth}
  \centering  \includegraphics[width=0.99\linewidth]{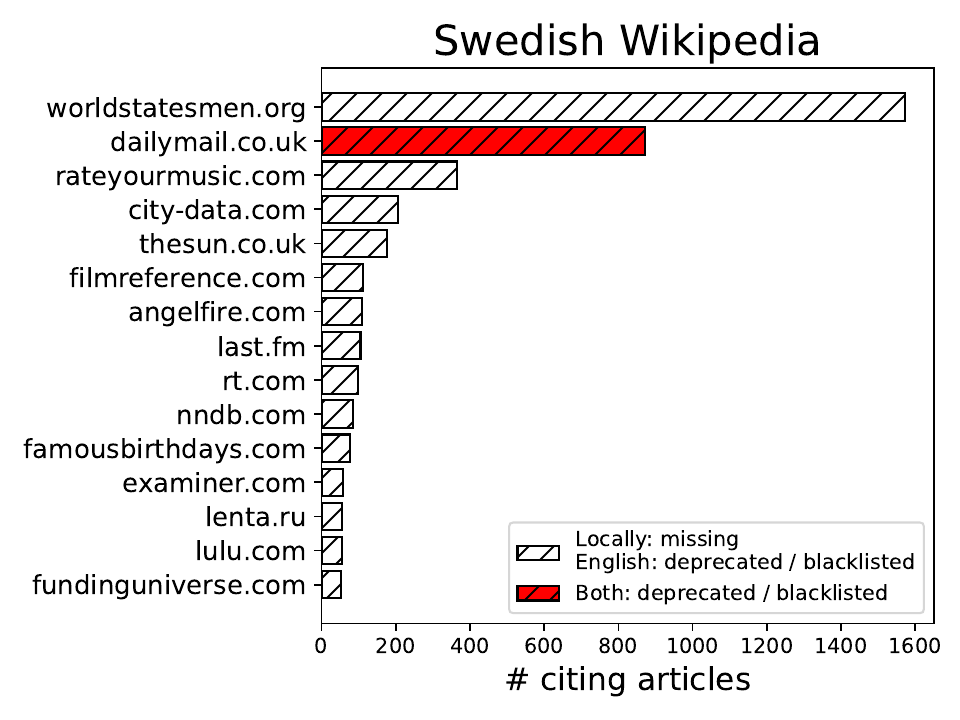}
  \label{fig:sv_barplot}
\end{subfigure}
\begin{subfigure}{.32\textwidth}
  \centering  \includegraphics[width=0.99\linewidth]{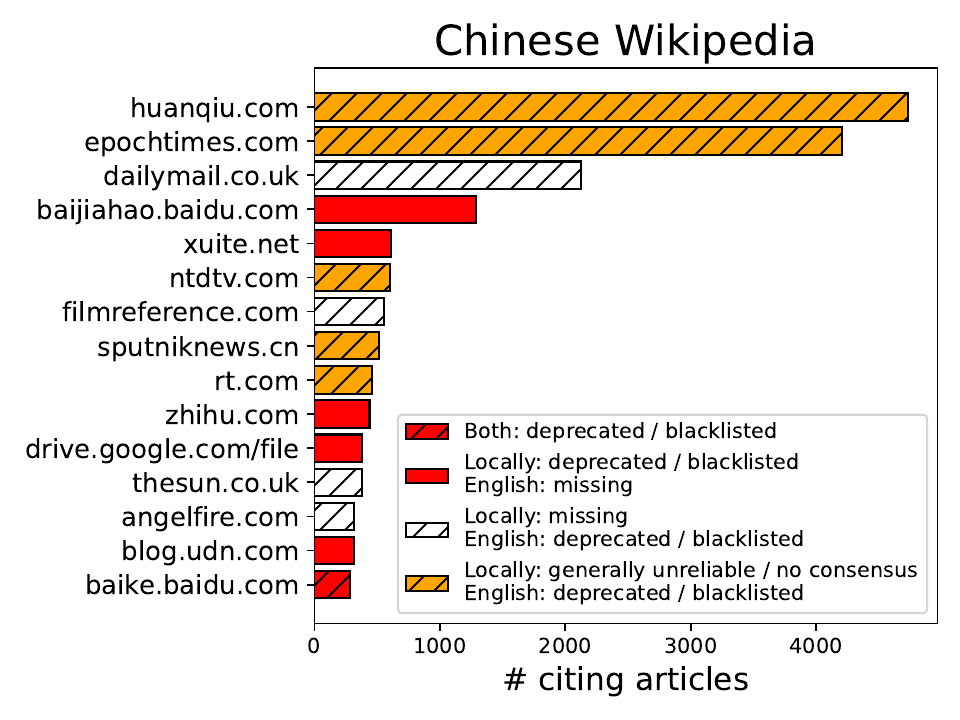}
  \label{fig:zh_barplot}
\end{subfigure}
\vspace{-5mm}
\caption{Top 15 non-authoritative sources (from the \textit{perennial source} list of the local Wikipedia edition or the one of English Wikipedia) by the number of citations in Russian, Swedish, and Chinese Wikipedia editions. }
\label{fig:barplots}
\end{figure*}

There may be various reasons why references in one language may reappear in another. One such method is translation. The proportion of articles created through translation among all pages that currently reference at least one non-authoritative domain is found among the top 50 editions with the most articles. While less than 5\% of these pages were created via translation in large language editions (e.g., Russian, Chinese, German, Italian, and Japanese), a significant proportion of pages with non-authoritative sources were created in this way in editions such as Uzbek (73\%) or Hebrew (58\%). These results suggest an important risk to content verifiability when translating articles into less developed Wikipedia language editions, as they have fewer resources to assess whether the original content included references to non-authoritative sources.

\subsection{Local Source Reliability Labeling} 
Finally, we compare the impact of internal initiatives that maintain reference reliability in smaller editions to those implemented on English Wikipedia. In particular, we analyze the \textit{perennial sources lists} from three language editions: Chinese, Swedish, and Russian. These lists are created by local editors through discussion and agreement, just like the English Wikipedia, making them comparable. Domains from the English list are frequently cited in articles written in all three languages. In the Russian edition, as shown in Table~\ref{tab:perennials}, domains in the English \textit{perennial sources list} (11.6\%) are cited in more than twice as many pages as in the Russian \textit{perennial sources list} (5.7\%) . 

Next, we investigate the non-authoritative domain category of \textit{perennial sources} for the three languages. Figure~\ref{fig:barplots} shows the top 15 most frequently appearing deprecated and blacklisted sources from the \textit{perennial source list} of the local language or English editions. A few domains are only included in the local lists (solid red bars). Some examples are \textit{news.rambler.ru} in Russian Wikipedia and \textit{xuite.com} in Chinese Wikipedia. Moreover, it is common that  domains marked as deprecated and blacklisted in English Wikipedia are missing in the local list (white dashed bars) or have not reached consensus in the Russian and Chinese editions (orange dashed bars).
\section{Conclusion} 
\noindent \textbf{Implications~} We presented a comparative study of reference reliability on Wikipedia across multiple language editions to understand the status quo of knowledge integrity~\cite{aragon2021preliminary}. We started from a recent analysis of the community initiative labeling domain reliability (called \textit{perennial sources list} in English Wikipedia)~\cite{ours} and examined their prevalence in other language editions.

Our finding that trustworthy sources considered by English editors are also more frequently cited in other language editions suggests that overall reliability is well maintained. On the other hand, non-authoritative domains that somehow persisted in the English edition also appeared in the same articles in other languages, indicating that reliability risks in a major language edition will percolate to smaller language editions. Because smaller language editions lack the same level of human resources to maintain each page, these risks may have spread through translation.

These findings suggest the potential that language editions could co-share and co-develop the \textit{perennial sources list}. This process will likely require global coordination among editors of various language editions, as well as agreement on what constitutes appropriate "global knowledge." Even in our case study, we found that cultural- or regime-specific domains were actively used in Wikipedia's local language edition, whereas the same domains were deemed unreliable in other language editions. One of the examples is a daily tabloid, \textit{huanqui.com} that supports the Chinese Communist Party. This domain frequently appears in the Chinese Wikipedia edition, yet it is labeled as untrustworthy in the English edition. Given the disparity in what constitutes reliable sources across language editions, coordination among Wikipedia editors on deciding ``global knowledge'' may be challenging. Nonetheless, policy decisions are needed because Wikipedia content has become a key database for search searches and training large language models.

There could be a possibility of relying on external domain ratings (for example, see~\cite{yang2022polarization}) rather than maintaining a Wiki-specific list. According to one study, external reliability ratings have a high level of agreement among experts~\cite{lin2022high}. However, studies also acknowledge the challenges of handling culturally specific biases. For example, biases in Wikipedia citations to scholarly publications have been discovered, favoring authors affiliated with Anglosphere countries~\cite{zheng2023gender}. Because Wikipedia source reliability labels must result from a deliberative process among community members, discussions must include smaller language editions, particularly for local sources unique to that language.

\vspace{1.5mm}
\noindent \textbf{Future Work~} 
We encountered several limitations that should encourage future work. First, we presented an analysis based on the most recent page versions without considering the articles' editing history. In Wikipedia, not only articles but also its rules evolve over time, including the decision to deprecate sources, as happened with the British tabloid \textit{dailymail.co.uk} after an intense debate~\cite{steinsson2023rule}. Further research could look into the temporal patterns of managing untrustworthy sources across languages, similar to previous work on reference quality in English Wikipedia~\cite{ours}.
Second, our analysis only considered the \textit{perennial sources list} in English Wikipedia. This is because comparable lists in other language editions are much less developed. Some editions had a higher percentage of articles referencing deprecated or blacklisted domains. Therefore, we emphasize the need for more active discussions within non-English communities, and we hope that our work inspires local initiatives to improve the reference quality of the content.

\vspace{1.5mm}
\noindent \textbf{Acknowledgements~} We thank the Wikipedia volunteers who contribute to content verifiability. This research was supported by the Institute for Basic Science (IBS-R029-C2) and the National Research Foundation of Korea (NRF) grant (RS-2022-00165347).


\bibliographystyle{ACM-Reference-Format}
\balance
\clearpage
\bibliography{reference}

\appendix
\end{document}